\documentclass[prb,showpacs,preprintnumbers,amsmath,amssymb,printfigures,twocolumn,superscriptaddress,10pt]{revtex4-1}
\usepackage{longtable}
\usepackage[colorlinks=true,allcolors=blue]{hyperref}
\usepackage{graphicx}
\usepackage{amsmath}
\usepackage{bm}
\usepackage[T1]{fontenc}
\usepackage[latin9]{inputenc}
\usepackage{textcomp}
\usepackage{gensymb}
\usepackage{amstext}
\usepackage[Symbol]{upgreek}
\usepackage{booktabs}
\usepackage{dcolumn}
\usepackage{color}
\usepackage{enumerate}
\usepackage{latexsym,bm}
\usepackage{setspace}
\usepackage{braket}
\usepackage{float}
\usepackage{array}
\usepackage{multirow}
\makeatletter

\begin{document}
\title{High-Throughput Computational Screening of Two-Dimensional Semiconductors}
\author{V. Wang}
\thanks{wangvei@icloud.com}
\affiliation{Department of Applied Physics, Xi'an University of Technology, Xi'an 710054, China}  

\author{G. Tang}
\affiliation{Advanced Research Institute of Multidisciplinary Science, Beijing Institute of Technology, Beijing 100081, China} 

\author{R. T. Wang}
\affiliation{Department of Applied Physics, Xi'an University of Technology, Xi'an 710054, China}  

\author{Y. C. Liu}
\affiliation{Department of Applied Physics, Xi'an University of Technology, Xi'an 710054, China} 

\author{H. Mizuseki}
\affiliation{Korea Institute of Science and Technology (KIST), Seoul 02792, Republic of Korea}  

\author{Y. Kawazoe}
\affiliation{New Industry Creation Hatchery Center, Tohoku University, Sendai, Miyagi 980-8579, Japan} 
\affiliation{Department of Physics and Nanotechnology, SRM Institute of Science and Technology, Kattankulathur,Tamil Nadu-603203, India} 
\affiliation{Department of Physics, Suranaree University of Technology, Nakhon, Ratchasima, Thailand} 

\author{J. Nara}
\affiliation{National Institute for Materials Science, Tsukuba 305-0044, Japan}

\author{W. T. Geng}
\thanks{geng@hainanu.edu.cn}
\affiliation{School of Materials Science and Engineering, Hainan University, Haikou 570228, China}

\date{\today}

\begin{abstract}

By performing high-throughput first-principles calculations combined with a semiempirical van der Waals dispersion correction, we have screened 74 direct- and 185 indirect-gap two dimensional (2D) nonmagnetic semiconductors from near 1000 monolayers according to the criteria for energetic, thermodynamic, mechanical, dynamic and thermal stabilities, and conductivity type.  We present the calculated lattice constants, simulated scanning tunnel microscopy, formation energy, Young's modulus, Poisson's ratio, shear modulus, anisotropic effective mass, band structure, band gap, ionization energy, and electron affinity for each candidate meeting our criteria. The resulting 2D semiconductor database (2DSdb) can be accessed via the website \url{https://materialsdb.cn/2dsdb/index.html}. The 2DSdb provides an ideal platform for computational modeling and design of new 2D semiconductors and heterostructures in photocatalysis, nanoscale devices, and other applications. Further, a linear fitting model was proposed to evaluate band gap, ionization energy and electron affinity of semiconductor from the density functional theory (DFT) calculated data as initial input. This model can be as precise as hybrid DFT but with much lower computational cost.

\end{abstract}
\maketitle

\section{Introduction}
Since the successful isolation of graphene,\cite{Novoselov2004,Novoselov2005} two dimensional (2D) materials have attracted tremendous attentions due to their novel electronic, optical, thermal, and mechanical properties for potential applications in a great variety of fields. Owing to the quantum confinement effect along the out-of plane direction, 2D materials often exhibit unique features, different from those of their bulk counterparts.\cite{Novoselov2005a,Schwierz2010,Song2012,Wang2012c,Zhang2005,Chhowalla2013,Xu2013,Butler2013,Zeng2015,Balendhran2015,Wang2015,Liu2015,Wang2015a} For examples, an unusual half-integer quantum Hall effect was observed in graphene.\cite{Zhang2005} The electronic properties of transition-metal dichalcogenides (TMDs) with MX$_2$ composition (where M = Mo or W and X = S, Se or Te) can be tuned from metallic to semiconducting by controlling layer-thickness.\cite{Mak2010,RadisavljevicB2011,Wang2012c,Chhowalla2013,Wang2015a,Manzeli2017} The peculiar puckered honeycomb structure of few-layer black phosphorene (BP) leads to significant anisotropic electronic and optical properties on zigzag and armchair directions.\cite{Li2014,Liu2014,Liu2015} Remarkably, its band gap is also thickness-dependent, varying from 0.3 eV in the bulk limit to $\sim$2.2 eV in a monolayer with a direct band gap character. 
Other 2D materials, such as hexagonal boron nitride ($h$-BN),\cite{Watanabe2004} silicene,\cite{Sugiyama2010,Okamoto2010,Hare2012,Vogt2012} germanene,\cite{Bianco2013,Li2014b} stanene,\cite{Zhu2015} also exhibit many exotic characteristics that are absent in their bulk form.

A common feature of 2D materials is that they are formed by stacking layers with strong in-plane bonds and weak van der Waals (vdW)-like interlayer attraction with typical binding energies of dozens of meV, allowing exfoliation into individual and atomically thin layers. This means that 2D materials usually possess in-plane stability in the absence of dangling bonds, in contrast to bulk films that are plagued by dangling bonds and surface state. Inspired by this feature, Inoshita \emph{et al.} screened the potential 2D binary stoichiometric electrides from the layered crystal structures by performing first-principles calculations based on the density functional theory (DFT) within the generalized gradient approximation (GGA).\cite{Inoshita2014} Later, Ahston and co-workers used a topology-scaling algorithm combining high-throughput calculations to uncover more than 800 monolayers based on the Materials Project crystal structure databases.\cite{Ashton2017,Jain2013} Considering the fact that the semi-local density functionals such as GGA functional significantly overestimates the lattice constants of crystals having vdW bonds. A rough thumb rule is that if the relative error in lattice constant $a$ or $b$ or $c$ (experimental versus GGA-calculated) of one bulk phase is larger than 5\%, it might have 2D structure.  Choudhary \emph{et al.} identified at least 1300 monolayers by comparing the experimental lattice constants with those predicted using the GGA functional.\cite{Choudhary2017} Cheon \emph{et al.} also identified thousand of 2D layered materials based on data mining algorithm.\cite{Cheon2017} Another important database for 2D materials was builded by Mounet \emph{et al}.\cite{Mounet2018} They chose the binding energy obtained by DFT calculations with the vdW correction as the screening criterion ($\leq$ few tens of meV$\cdot${\AA}$^{-1}$) and identified more than 1800 structures.
There are several 2D crystals databases publicly available at present, such as MC2D,\cite{Mounet2018} C2DB\cite{Haastrup2018}, 2DMatPedia\cite{Zhou2019} and JARVIS-DFT\cite{Choudhary2017}. However, one of the major limitations of these databases is that they mainly focus on the stability analysis and provide only a small number of the fundamental physical properties such as lattice constants, formation energy, exfoliation energy, and band gap at the GGA level. Although the GGA functional can provide sufficiently accurate results on forces, structures, and band dispersions, it underestimates band gap of semiconductors, averagely by 50\%. Furthermore, to our knowledge, the computational materials databases targeted on 2D semiconductor are still incomplete and a high-throughput screening of 2D semiconductors is strongly called for.

In this work, combined the high-throughput first-principles calculations with a semiempirical vdW dispersion correction, we have chosen the energetic, thermodynamic, mechanical, dynamic, thermal stabilities and conductivity type as the criteria and screened around 260 2D semiconductors from near 1000 structures. We here present the lattice constants, formation energy, simulated scanning tunnel microscopy (STM), Young's modulus, Poisson's ratio, phonon dispersions, band structure, effective masses of carriers, band gap, ionization energy and electron affinity for each candidate. The remainder of this paper is organized as follows. In Sec. II, methodology and computational details are described. The details of screening criteria are discussed in Sec. III. Sec. IV presents the calculations of structural, mechanical and electronic properties. Finally, a short summary is given in Sec. V.

\section{Methodology}
\subsection{Density Functional Calculations}

Our total energy calculations were performed using the Vienna Ab initio Simulation Package (VASP).\cite{Kresse1996, Kresse1996a} The electron-ion interaction was described using projector augmented wave (PAW) method \cite{PAW, Kresse1999} and the exchange and correlation (XC) were treated with GGA in the Perdew Burke Ernzerhof (PBE) form\cite{Perdew1996}. Part of electronic structure calculations were also performed using the standard screening parameter of Heyd-Scuseria-Ernzerhof (HSE06) hybrid functional, \cite{Becke1993,Heyd2003,Perdew1996a,Paier2006,Krukau2006,Marsman2008}  upon the PBE-calculated equilibrium geometries. A cutoff energy of 400 eV was adopted for the plane wave basis set, which yields total energy convergence better than 1 meV/atom. In addition, the non-bonding vdW interaction is incorporated by employing a semi-empirical correction scheme of Grimme's DFT-D2 method in this study, which has been successful in describing the geometries of various layered materials.\cite{Grimme2006, Bucko2010}  In the slab model of 2D systems, periodic slabs were separated by a vacuum layer of 20 {\AA} in \emph{z} direction to avoid mirror interactions. 
The Brillouin zone was sampled by the  \emph{k}-point mesh following the Monkhorst-Pack scheme,\cite{Monkhorst1976}  with a reciprocal space resolution of 2$\pi$$\times$0.03 {\AA}$^{-1}$. On geometry optimization, both the shapes and internal structural parameters of pristine unit-cells were fully relaxed until the residual force on each atom is less than 0.01 eV/{\AA}.  To screen the novel 2D semiconductors, we used the VASPKIT package\cite{vaspkit} as a high-throughput interface to pre-process the input files and post-process the data obtained by using VASP code. 

\subsection{High-Throughput Settings}
\begin{figure}[htbp]
\centering
\includegraphics[scale=0.46]{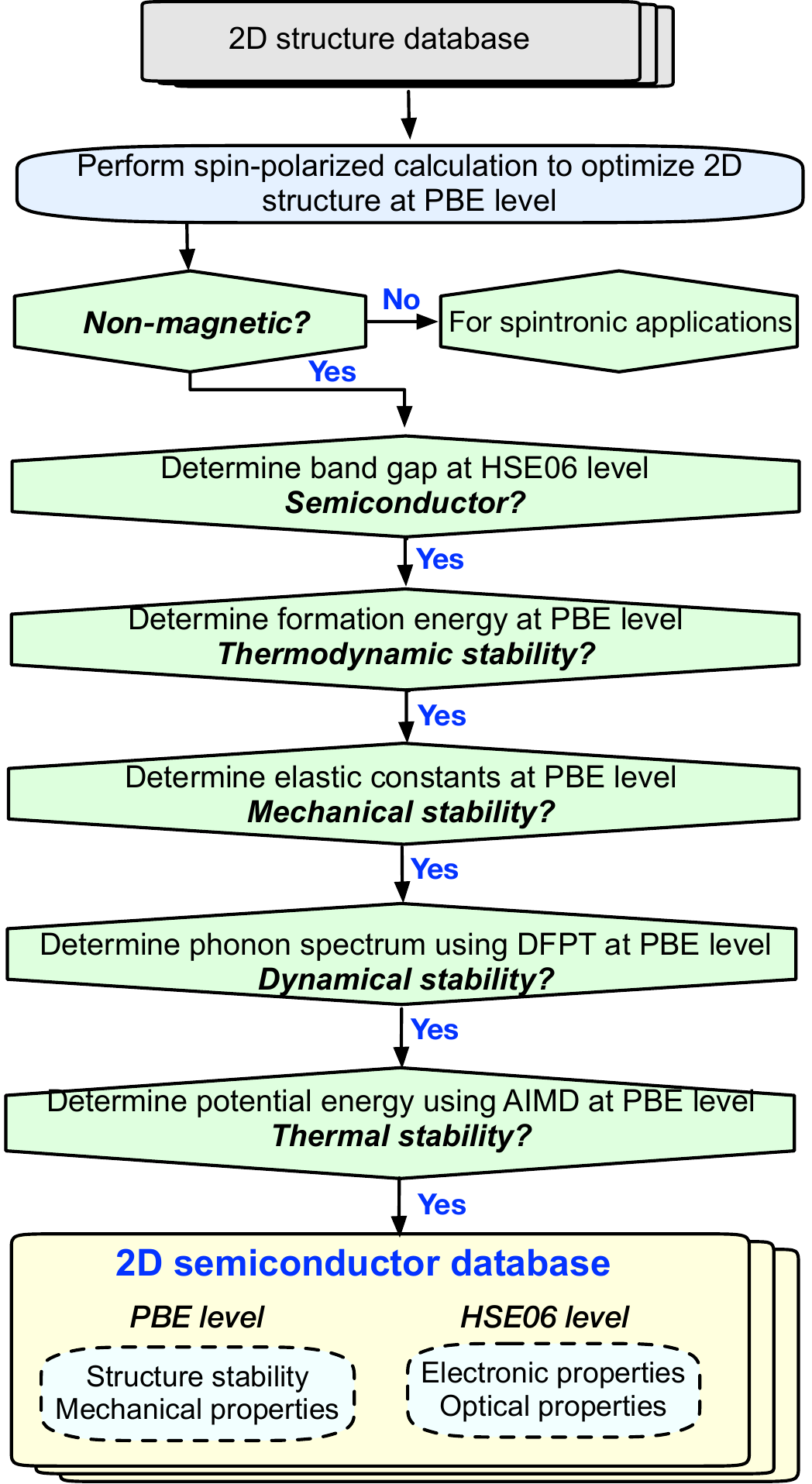}
\caption{\label{flow_chart}(Color online) Schematic representation of the fundamental steps needed to find 2D semiconductors and vdWHs photocatalysts.}
\end{figure}

The purpose of this work is to identify the candidates of 2D semiconductors and vdWHs photocatalysts through large-scale screening of 2D materials, rather than to make the most accurate prediction of a specific material.  To screen the novel 2D semiconductors, we used the VASPKIT package\cite{vaspkit} as a high-throughput interface to pre-process the input files and post-process the data obtained by using VASP code. The overview of the screening process is shown schematically in Fig. \ref{flow_chart}. First, VASPKIT generates three input files (POTCAR, KPOINTS, and INCAR) for a given structure file (POSCAR). Then the spin-polarized structure-relaxation was done to determine the magnetic ground state for each 2D material. If the candidate is nonmagnetic, we next calculated the global band structure at the PBE level to determine the accurate positions of both conduction-band minimum (CBM) and valence-band maximum (VBM) in the reciprocal space. It is well known that the PBE functional is sufficiently accurate on band dispersion, but underestimates band gaps. The HSE06 can well describe narrow and middle-sized gap semiconductors whose valence electrons are not strongly localized.\cite{Heyd2005,Krukau2006} 
Thus, the band structure calculations at HSE06 level were performed in order to get accurate band gap $E_g$ values at the PBE-calculated lattice constants. If the candidate meets the energetic, thermodynamic, mechanical, dynamic and thermal stability criteria and bears a non-zero band gap, it could be a potential 2D semiconductor. Finally, we have further screened potential 2D semiconductors and vdWHs photocatalysts according to the photocatalytic criteria which willl be discussed later. This screening algorithm is expected to be applicable to other fields, such as 2D thermoelectricity materials.

\section{Screening Criteria}
\subsection{Thermodynamic Stability}
Generally speaking, a stable material should have thermodynamic, mechanical, dynamic and thermal stabilities simultaneously.  Thermodynamic stability measures the steadiness of a compound against its decomposition. Three physical quantities are commonly used to evaluate the thermodynamic stability of a free-standing 2D sheet, namely, the exfoliation energy, the energy convex hull and the formation energy. The exfoliation energy is the energy needed to exfoliate a monolayer from its bulk, an indication of the strength of interlayer bonds holding the layered bulk structure together. However,  some 2D materials, such as borophene,\cite{Mannix2015} lack any layered bulk structure from which they can be exfoliated. The energy convex hull describes the competition between all phases with the same composition. Specifically, the phases lying  above the convex hull have a tendency to decompose into the ground state compounds on the convex hull. The definition of energy convex hull, nevertheless, has the same problem as the exfoliation energy.\cite{Paul2017,Malyi2019} For example, the synthesis of 2D sheets by mechanical exfoliation implies that it is an endothermic process to break the interlayer bonds. This means that all 2D materials with respect to their corresponding bulk counterparts naturally fall above the convex hulls. The formation energy which is defined as the difference between a material and its pure elemental constituents in their ground states
\begin{equation}\label{eq1}
\Delta E_f=E_{tot}-\sum n_{\alpha}\mu_{\alpha},
\end{equation}
where $E_{tot}$ is the total energy of pristine 2D monolayer. \emph{n}$_\alpha$ is the number of atoms of species $\alpha$ and $\mu_{\alpha}$ is the atomic chemical potential of species $\alpha$ which is equal to the total energy of per atom in its most stable elemental phase. A more negative $\Delta E_f$ for a material means higher thermodynamic stability. However, to be thermodynamically stable, a material must not only have a negative formation energy not only with respect to the elemental ground states but also have a negative one with respect to all possible competing compound  phases. 
In the present study we mainly focus on the high-throughput computational screening of 2D semiconductors, and adopt the PBE-calculated formation energy as the thermodynamic stability criteria. PBE generally underestimate the formation energy of solids, especially for the layered materials, with an accuracy of  only around 0.2 eV/atom on average.\cite{Pandey2015}  We noted that the PBE-calculated formation energies of Si, Ge and Sn monolayer are higher than 0.6 eV/formula-unit (f.u.) but they have recently been synthesized or isolated by exfoliation.\cite{Vogt2012,Bianco2013,Zhu2015} In our high-throughput screening process, we used a threshold of 1.0 eV/f.u. as an upper bound on the thermodynamic stability for free-standing monolayers. 

\subsection{Mechanical Stability}
The mechanical stability of a material describes its resistance to deformations or distortions in the presence of strain.
For a 2D crystal in the linear elastic region, the stress $\mathbf{\sigma} = \left(\sigma_{1}, \sigma_{2}, \sigma_{6}\right)$ response to external loading strain $\boldsymbol{\varepsilon} = \left(\varepsilon_{1}, \varepsilon_{2}, \varepsilon_{6}\right)$ follows the generalized Hooke's law and can be simplified in the Voigt notation, \cite{Voigt1928,Mazdziarz2019}

$\left(\begin{array}{c}{\sigma_{1}} \\ {\sigma_{2}} \\ {\sigma_{3}}\end{array}\right)=\left(\begin{array}{ccc}{\text{C}_{11}} & {\text{C}_{12}} & {\text{C}_{16}} \\ {\text{C}_{21}} & {\text{C}_{22}} & {\text{C}_{26}} \\ {\text{C}_{61}} & {\text{C}_{62}} & {\text{C}_{66}}\end{array}\right) \cdot\left(\begin{array}{c}{\varepsilon_{1}} \\ {\varepsilon_{2}} \\ {\varepsilon_{6}}\end{array}\right),$

where \emph{C}$_{ij}$ (\emph{i},\emph{j}=1,2,6) is the in-plane stiffness tensor using the standard Voigt notation: 1-\emph{xx}, 2-\emph{yy}, and 6-\emph{xy}. The $\emph{C}_{ij}$ can be obtained using the energy-strain method as outlined in our previous computational study on the mechanical anisotropy of borophene,\cite{Wang2017} namely, 

\begin{equation}\label{eq1}
\begin{split}
E_{\text {elastic}}\left(E,\left\{\varepsilon_{i}\right\}\right)=E(S,&\left.\left\{\varepsilon_{i}\right\}\right)-E\left(S_{0}, 0\right) \\ 
 =\frac{S_0}{2} (C_{11} \varepsilon_{1}^{2}+ \mathrm{C}_{22} \varepsilon_{2}^{2}+2C_{12} \varepsilon_{1} \varepsilon_{2} \\ +  2C_{16} \varepsilon_{1} \varepsilon_{6}+2C_{26} \varepsilon_{2}\varepsilon_{6}+ C_{66} \varepsilon_{6}^{2}).
\end{split}
\end{equation}

\begin{figure*}[htbp]
\centering
\includegraphics[scale=0.92]{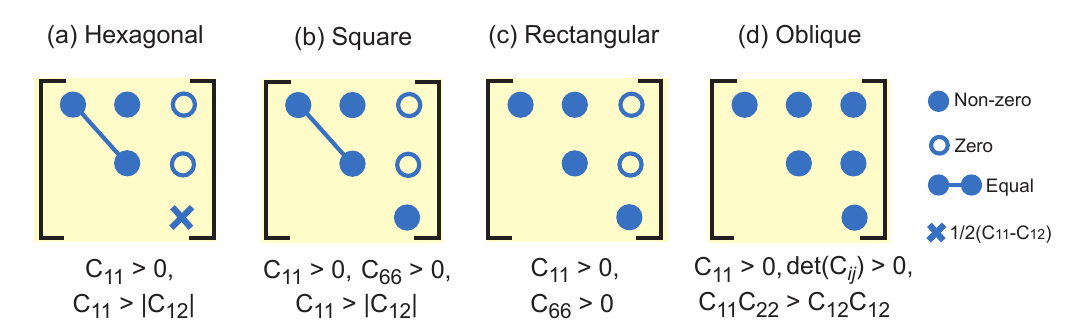}
\caption{\label{elastic_2D} Classification of crystal systems, independent elastic constants, elastic stability conditions for 2D materials.}
\end{figure*}

In the energy-strain method, the C$_{ij}$ is equal to the second partial derivative of strain energy $E_{elastic}$ with respect to strain $\varepsilon$, and can be written as C$_{ij}=(1/S_{0})(\partial^{2}E_{elastic}/\partial\varepsilon_{i}\partial\varepsilon_{j})$, where $S_{0}$ is the equilibrium area of the system. Therefore, the unit of elastic stiffness constants for 2D materials is force per unit length (N/m).
In order to calculate C$_{ij}$, the $E_{elastic}$ as a function of $\varepsilon$ in the strain range -2\% $\leq \varepsilon \leq$ 2\% with an increment of 0.5\% were investigated. The number of independent elastic constants is controlled by the symmetry of a 2D crystal. For instance, the hexagonal crystals have two but the oblique ones have six independent elastic constants. This number,  together with the necessary and sufficient elastic stability conditions for different 2D lattice types are summarized in Fig. \ref{elastic_2D}.\cite{Born1954,Mazdziarz2019}

\subsection{Dynamic Stability}
\begin{figure*}[htbp]
\centering
\includegraphics[scale=0.65]{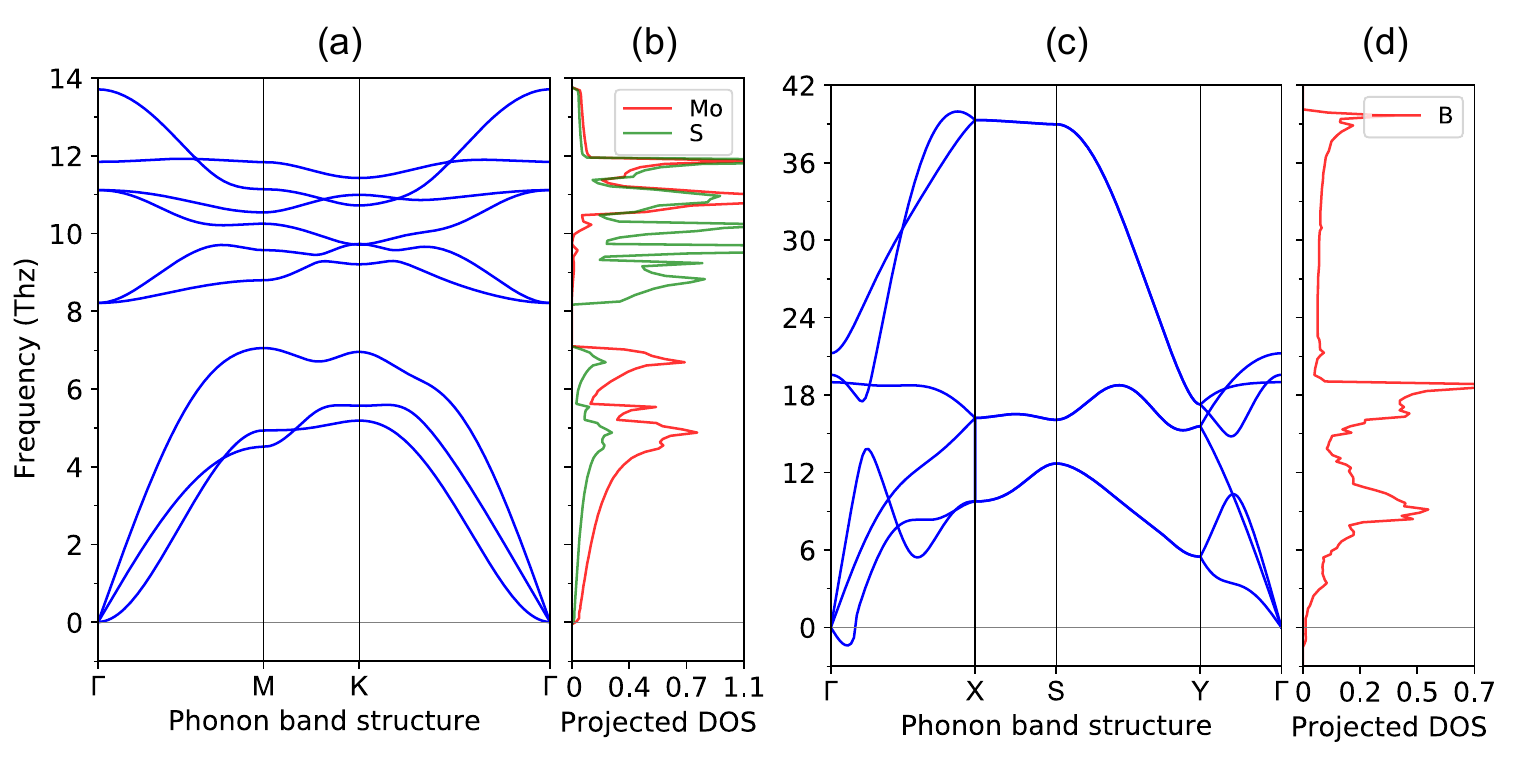}
\caption{\label{phonon}(Color online) Calculated (a) and (c) phonon dispersion curves, projected density of states (b) and (d) for H-MoS$_2$ and borophene sheet respectively.}
\end{figure*}

The dynamic stability reflects the structural tolerance of a system against small atomic displacements due tothermal motions. It can be determined by calculating the phonon dispersions of a material using either a finite displacement method \cite{Parlinski1997} or density functional perturbation theory\cite{Baroni2001}. We derived phonon dispersions using the finite displacement approach implemented in the PHONOPY code.\cite{Togo2015}  The force constants were calculated using a supercell (20 {\AA} $\times$ 20 {\AA}) with atomic displacements of 0.01 {\AA} along the lattice vectors. To be dynamically stable, a material allows no imaginary phonon spectra in its phonon dispersions. Shown in Fig. \ref{phonon} (a) is the phonon spectra of hexagonal MoS$_2$ monolayer. No imaginary modes appear, implying that is dynamically stable. Otherwise, the material will undergo reconstructive or martensitic phase transformations upon a slight lattice distortion.

It is worth mentioning that small negative spectra, $\emph{i.e.}$, low imaginary frequency near the $\Gamma$ point is often observed in the phonon spectra of 2D systems, as is the case for borophene monolayer [Fig. \ref{phonon} (c)] which has been synthesized recently.\cite{Mannix2015} Such small imaginary frequencies could be an artifact of poor convergence due to limited supercell size, cutoff energy, or $k$-points; or they may reflect the actual lattice dynamical instability towards large wave undulations of 2D materials. It can possibly be eliminated by applying a small strain on the film or depositing the film onto a proper substrate.\cite{Mannix2015,Penev2016} Thus, a candidate is still considered to be dynamically stable even if a tiny imaginary frequency is present near the $\Gamma$ point. We note that the phonon criterion is still a necessary but not sufficient condition to evince dynamic stability of a material. Since the phonon analysis deals only with small atomic displacements, it cannot capture phase transitions coupled with complex lattice reconstructions.\cite{Malyi2019}

\subsection{Thermal Stability}
\begin{figure*}[htbp]
\centering
\includegraphics[scale=0.48]{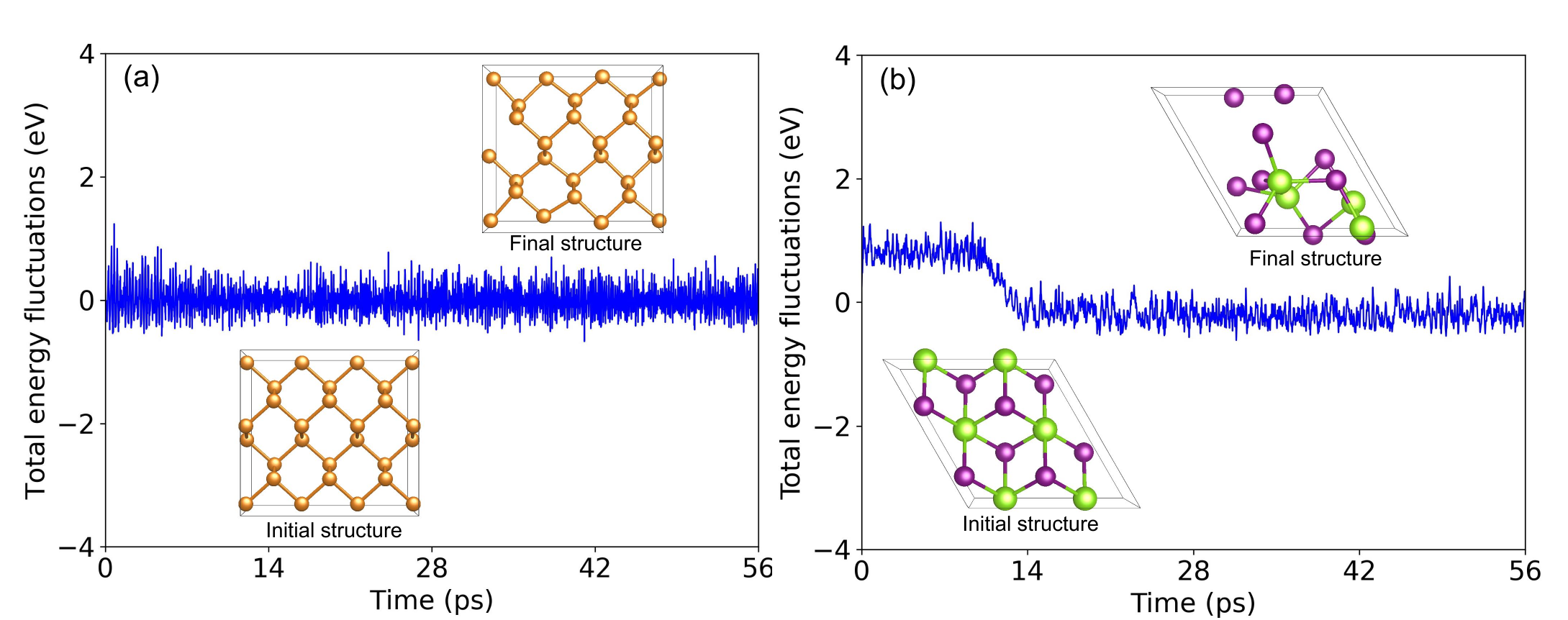}
\caption{\label{md}(Color online) Total potential energy fluctuations of (a) BP ($Pmma$)  and (b) MgI$_2$ ($P\overline{3}m1$) during AIMD simulations at 400 K. The inset shows the snapshots at the begin and end of simulation. The results show that MgI$_2$ tends to reconstruct into lower energy structure and is unlikely to be realized experimentally in the freestanding forms. }
\end{figure*}

Finally, the thermal stability of a material reflects its resistance to decomposition or reconstruction into lower energy structures at high temperatures, and can be evaluated by performing \emph{ab-initio} molecular dynamics (AIMD) simulations  over a long time and wide range of temperatures.  To verify the dynamic stability of the proposed 2D materials, we employed AIMD simulations of a 10 {\AA} $\times$ 10 {\AA} supercell model at a temperature of 400 K. The  time step and time duration are set to 1.0 fs and 60 ps, respectively. A Nos{\'e}-Hoover thermostat was used to control the temperature.\cite{Martyna1992}  To be dynamically stable, its potential energy should remain roughly constant during the AIMD simulation. For comparative purpose, we found that the calculated potential energy of BP ($Pmma$) fluctuates around the equilibrium state as a function of time [Fig. \ref{md} (a)], indicating a good thermal stability. In contrast, the potential energy of MgI$_2$ ($P\overline{3}m1$) decreases over time, reflecting an irreversible change in structure which lowers the formation energy. The snapshot of its atomic configuration at the end of the simulation further shows that  this material is  drastically distorted and is unlikely to be fabricated in the free-standing forms. 

\subsection{Semiconductor Screening}
\begin{figure*}[htbp]
\centering
\includegraphics[scale=0.55]{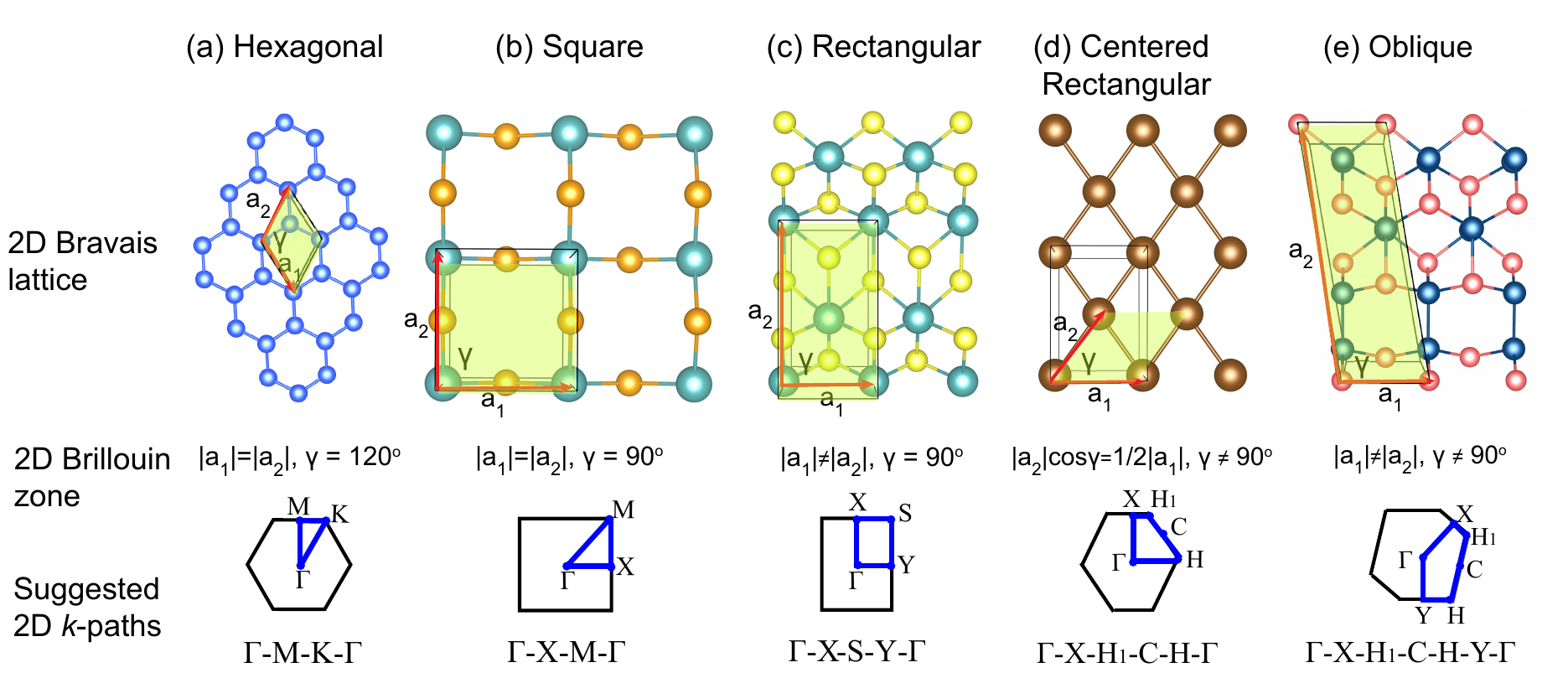}
\caption{\label{fig_kpath}(Color online) Overview of the five 2D Bravais lattices and corresponding Brillouin zones. The suggested $k$-paths for band structure are indicated in blue line. The primitive unit cell is indicated in green box.}1/3
\end{figure*}

For nonmagnetic semiconductors, the Kohn-Sham (KS) band gap $E_g$  is defined as the difference between the eigenvalues of CBM and VBM. That is, 

\begin{equation}\label{eq2}
E_g=\epsilon_\text{CBM}-\epsilon_\text{VBM},
\end{equation}

where $\epsilon_\text{CBM}$ and $\epsilon_\text{VBM}$ are the KS eigenvalues of CBM and VBM respectively. It is well known that PBE severely underestimates the band gap of semiconductors  because of the lack of derivative discontinuity of the functional with respect to the number of electrons and the lack of  clear physical meaning of the unoccupied orbitals. But  PBE yields similar band dispersion curves to the hybrid DFT result. 
There are five typical 2D Bravais lattices, namely, hexagonal, square, rectangular, centered rectangular, and oblique respectively. The Ball-and-stick models, Brillouin zones and suggested $k$-paths for the Bravais lattices adopted in our high-throughput calculations are presented in  Fig. \ref{fig_kpath}  and Table \ref{table_kpath}.

\begin{table*}[htbp]
\begin{ruledtabular}
\caption{\label{table_kpath} Fractional coordinates of the specific points in reciprocal space for the four nonequivalent two-dimensional Bravais lattices. }
\begin{tabular}{c|c|c|c|}
Bravais Lattice & Label and coordinates of specific points & Bravais Lattice & Label and coordinates of specific points   \\
\hline
 & $\pmb{\Gamma}$ (0, 0)  &    & $\pmb{\Gamma}$ (0, 0)  \\
Square & \textbf{X} (1/2, 0) &  Oblique  &  \textbf{X} (1/2, 0)       \\
       & \textbf{M} (1/2, 1/2)  &   & \textbf{Y} (0, 1/2)    \\
 \hline
 &$\pmb{\Gamma}$ (0, 0) &  & \textbf{C} (1/2, 1/2)     \\
Hexagonal   & \textbf{K} (1/3, 1/3)   &  Oblique  &  \textbf{H} ($\eta$, 1-$\nu$)$^a$  \\
 & \textbf{M} (1/2, 0) &   & \textbf{H$_1$} (1-$\eta$, $\nu$)$^a$   \\
 \hline
Rectangular & $\pmb{\Gamma}$ (0, 0)  &  Rectangular   & \textbf{X} (1/2, 0)        \\
 &\textbf{Y} (0, 1/2) &     & \textbf{S} (1/2, 1/2)   \\
\end{tabular} 
\leftline{$^a$ $\eta=\frac{1-acos\gamma/b}{2sin^{2}\gamma}$, $\nu=\frac{1}{2}-\frac{\eta bcos\gamma}{a}$ and $\gamma<90\degree$.}
\end{ruledtabular}
\end{table*}

\section{Results and Discussions} 

\begin{figure*}[htbp]
\centering
\includegraphics[scale=0.66]{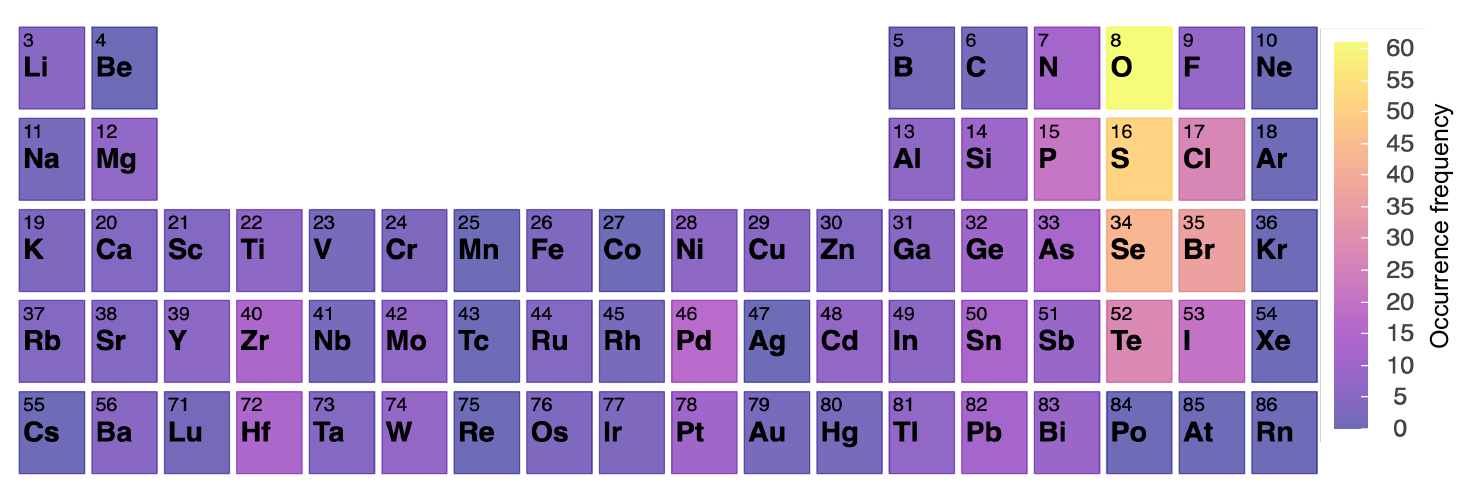}
\caption{\label{ptable}(Color online) Heat map of the occurrence frequency of each element in the screened 2D semiconductors.}
\end{figure*}

\begin{figure*}[htbp]
\centering
\includegraphics[scale=0.25]{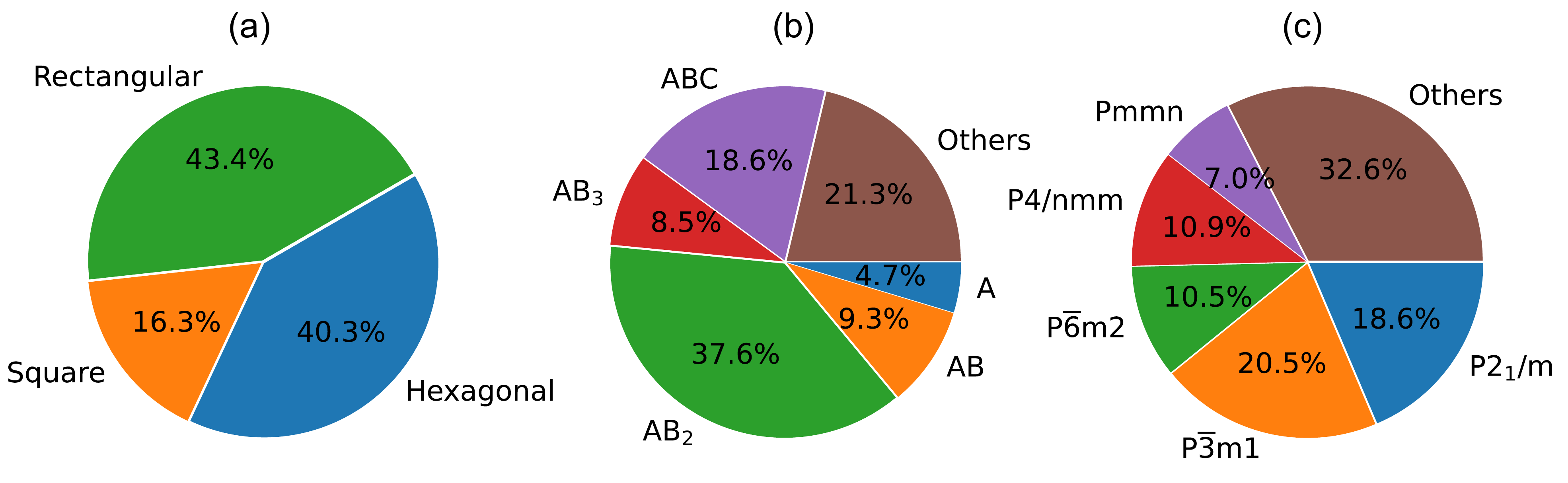}
\caption{\label{classification}(Color online) Classification of the screened 2D semiconductors in term of (a) lattice type, (b) stoichiometry and  (c) symmetry.}
\end{figure*}

Based on the above criteria, we have screened 74 direct- and 185 indirect-gap 2D nonmagnetic semiconductors from near 1000 2D monolayers. By analyzing the occurrence frequency of each element in the screened 2D semiconductors shown in Fig. \ref{ptable}, it is found that the most abundant candidates are oxides, followed by sulfides, selenides and  halides. Meanwhile, the cations appear to favor heavy metal elements such as Pd, Zr, Hf and Pb. The classifications of these candidates according to the relative frequencies of lattice type, stoichiometry and space group of the crystals are further summarized in Figs. \ref{classification}(a)-(c), respectively. Note that the lattice types of 2D semiconductors are dominated by rectangular (43.4\%) and hexagonal (40.3\%), and the least abundant are square (16.3 \%). Most of them are binary compounds predominantly bearing by AB$_2$ structures. Moreover, the space groups of these candidates are mainly P2$_1$/m and P$\overline{3}$m1. It is noteworthy that TMDs are one of the most interesting families in the AB$_2$ layered compounds and display a wide range of important properties. The TMD monolayers have three phases, namely, 2H (P$\overline{6} m 2$), 1T (P$\overline{3} m 1$) and 1T' (P2$_1$/m), respectively. Previous theoretical studies have predicted that around 50 different transition-metal oxides (TMOs) and TMDs can remain stable as either 2H and/or 1T free-standing structures,\cite{Ataca2012,Rasmussen2015} even though part of these potential MX$_2$ compounds are absent in their bulk counterparts. For the sake of completeness, we also revisited the stability and electronic structure of TMOs and TMDs with three possible phases (2H, 1T and 1T' respectively). We find that the band gap of these semiconducting candidates is mainly concentrated between 1.0 and 3.0 eV.  The structural, mechanical and electronic properties for each candidate are summarized in the Supplemental Material. 

\subsection{Mechanical Properties} 

\begin{figure*}[htbp]
\centering
\includegraphics[scale=0.75]{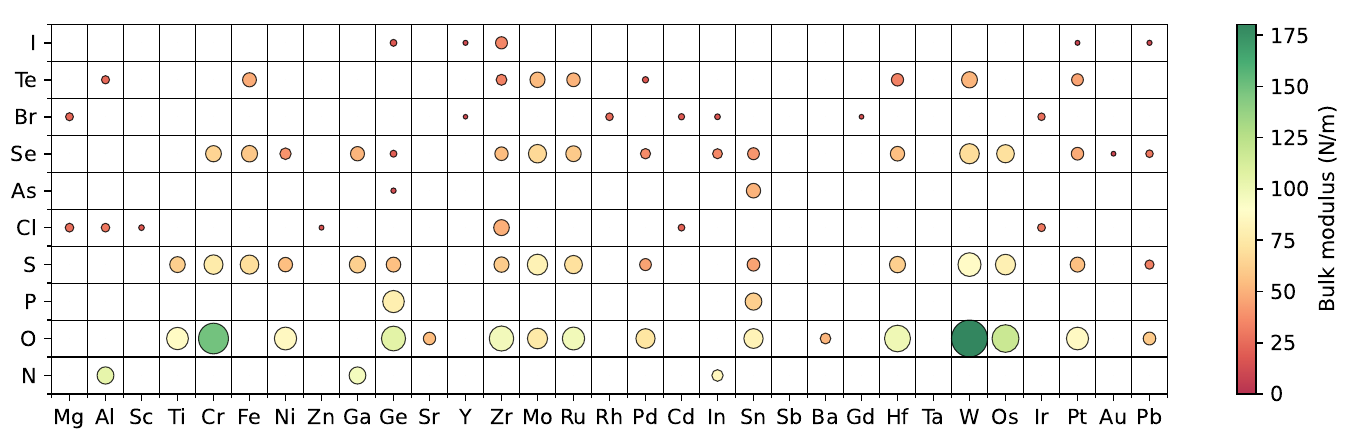}
\caption{\label{binary_modulus}(Color online) Bulk and shear modulus of binary 2D semiconductors as a function of  the constituent elements within Voigt-Reuss-Hill (VRH) approximation. The circle radius represents the magnitude of shear modulus.}
\end{figure*}

The mechanical properties of a single-crystal are generally anisotropict. The Voigt-Reuss-Hill (VRH) approximation,\cite{Li2020,Hill1952} is a useful scheme by which one can calculate isotropic polycrystalline elastic moduli in terms of the anisotropic single-crystal elastic constants. We present the VRH averaged bulk and shear moduli of binary 2D semiconductors as a function of the constituent elements in Fig. \ref{binary_modulus}. One can find that oxides have the largest bulk modulus, followed by sulfides and then selenides. As expected, the shear modulus indicates positive correlations with bulk modulus.
Next we compare our predicted data with available experimental or theoretical reports. Up until now, several monolayers have been successfully exfoliated or synthesized, including graphene (P6/mmm),\cite{Novoselov2004} BP (Pmna),\cite{Li2014,Liu2014,Liu2015} borophene (Cmmm),\cite{Mannix2015} BN (P$\overline{6} m 2$),\cite{Watanabe2004,Pacile2008} MoS$_2$ (P$\overline{6} m 2$),\cite{Mak2010} TiS$_3$ (P2$_1$/m)\cite{Island2015}. We summarize the calculated in-plane elastic stiffness constants, the minimum and maximum of Youngs's modulus, shear modulus and Poisson's ratio for these systems in Table \ref{tab_elastic}. One can find that our predictions  are in good agreement with the available published data. For example, the PBE-calculated Young's modulus and Poisson's ratio of graphene are 339 N/m and 0.17, in excellent agreement with the available values of 340 N/m and 0.186,\cite{Lee2008,Liu2007} respectively. To investigate the anisotropic mechanical properties of 2D materials, we also calculated the orientation-dependent Young's moduli $Y(\theta)$, Poisson's ratio $\nu(\theta)$ and shear modulus $G(\theta)$ using the following formulae,\cite{Jasiukiewicz2008,Jasiukiewicz2010a}

\begin{figure*}[htbp]
\centering
\includegraphics[scale=0.78]{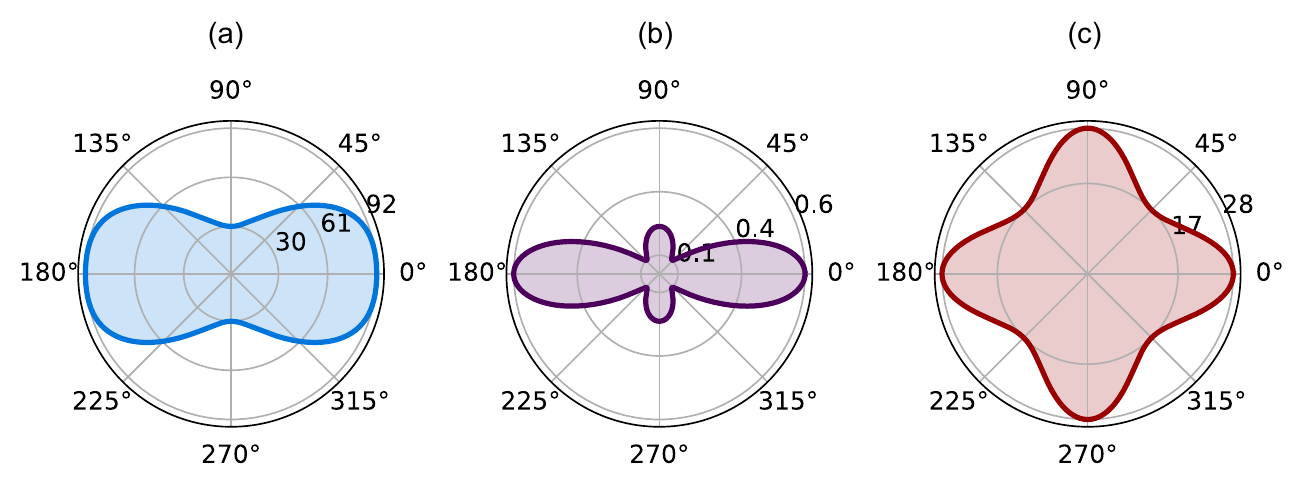}
\caption{\label{elastic_borophene}(Color online) Calculated orientation-dependent (a) Youngs's modulus $E(\theta)$, (b) Poisson's ratio $\nu(\theta)$ and (c) shear modulus $G(\theta)$ for BP respectively.}
\end{figure*}

\begin{equation}
\begin{aligned} 1 / E(\theta) =S_{11} c^{4}+S_{22} s^{4}+2 S_{16} c^{3} s \\+2 S_{26} c s^{3}+\left(S_{66}+2 S_{12}\right) c^{2} s^{2} \end{aligned},
\end{equation}

\begin{equation}
\begin{aligned} \nu(\theta) / E(\theta) =\left(S_{66}-S_{11}-S_{22}\right) c^{2} s^{2} \\-S_{12}\left(c^{4}+s^{4}\right)+\left(S_{26}-S_{16}\right)\left(c s^{3}-c^{3} s\right) \end{aligned},
\end{equation}

and

\begin{equation}
\begin{array}{l}{1 / 4 G(\theta)=\left(S_{11}+S_{22}-2 S_{12}\right) c^{2} s^{2}+} \\ {S_{66}\left(c^{2}-s^{2}\right)^{2} / 4-\left(S_{16}-S_{26}\right)\left(c^{3} s-c s^{3}\right)}\end{array},
\end{equation}

where $s=sin(\theta)$, $c=cos(\theta)$, and  $\theta$ $\in[0,2 \pi]$ is the angle with respect to the +\emph{x} axis. $\emph{S}_{ij}$= $\emph{C}_{ij}^{-1}$ are elastic compliance constants. As an example, It is found in Fig. \ref{elastic_borophene} that the mechanical properties of a BP monolayer shows a strong anisotropy. It is expected that all but hexagonal 2D bravais lattices have the anisotropic mechanical properties.

\begin{table*}[htbp]
\begin{ruledtabular}
\caption{\label{tab_elastic} PBE-calculated in-plane elastic stiffness constants, Youngs's modulus $Y(\theta)$, shear modulus $G(\theta)$ (in units of N/m), and Poisson's ratio $\nu(\theta)$. For comparison purposes, the available theoretical or experimental values from the previous literature are also shown.} 
\begin{tabular}{c|cc|cc|cc|cc|cc|cc|}
&\multicolumn{2}{c|}{C$_{11}$ }
&\multicolumn{2}{c|}{C$_{22}$}
&\multicolumn{2}{c|}{C$_{12}$ }
&\multicolumn{2}{c|}{E($\varphi$)}
&\multicolumn{2}{c|}{G($\varphi$)}
&\multicolumn{2}{c|}{$\nu(\varphi)$}\\
Systems & Calc. &  Refs. & Calc. & Refs.  & Calc. & Refs. & Max & Min & Max & Min & Max & Min    \\
\hline
Graphene & 349    & 342  [\onlinecite{Falin2017}]  & 349  & 342 [\onlinecite{Falin2017}] & 60 & -   & 339 & 339 & 144 & 144 & 0.17 & 0.17   \\
BP & 106    & 105  [\onlinecite{Wang2015c}]  & 34  & 26 [\onlinecite{Wang2015c}] & 22 & 18 [\onlinecite{Wang2015c}] & 92 & 29 & 28 & 17 & 0.63 & 0.08      \\
BN & 292    & 289 [\onlinecite{Falin2017}] & 292 & 289 [\onlinecite{Falin2017}] & 64   &  - & 277 & 277  & 114 & 114 & 0.22 & 0.22   \\
MoS$_2$  & 131    & 124 [\onlinecite{Kang2015}] & 131 & 124 [\onlinecite{Kang2015}] & 33 &-  & 122 & 122 & 49 & 49 & 0.26 & 0.26  \\
TiS$_3$  & 88    & 83 [\onlinecite{Kang2015}] & 137 & 134 [\onlinecite{Kang2015}] & 14   & - & 137 & 71 & 47 & 25 & 0.42 & 0.10   \\
\end{tabular} 
\end{ruledtabular}
\end{table*}
 
Thermodynamic stability sets limits on the energy and  the range of Poisson's ratio is allowed to be from -1.0 to 0.5. Most materials have a positive Poisson's ratio, shrinking (expanding) longitudinally after being stretched (compressed) laterally. We do find a few materials with a negative Poisson's ratio (NPR), also called auxetic materials. The NPR behavior is mainly attributed to some special re-entrant or hinged geometric structures regardless of the chemical composition and electronic structure of a material. The NPR materials exhibit fascinating mechanical properties, such as superior toughness, higher indentation resistance, larger impact resistance, stronger sound absorption, and better crack propagation resistance.\cite{Yang2004} These excellent properties offers enormous potential in many important applications, such as automotive, aerospace, marine, and other industrial fields.\cite{Greaves2011,Huang2016} Recently, the auxetic effect has been reported in a number of 2D materials. In addition to monolayer phosphorus and arsenic allotrope reported in previous studies,\cite{Jiang2014,Du2016,Han2015} we also screened some other 2D semiconductors with large NPR values, including As$_2$SO$_6$ (-0.392), SiP$_2$ (-0.320), BaIF (-0.256), GeSe (-0.228), SnS (-0.189) and SbSeI (-0.166). Among them, BaIF is the only one persisting the NPR in all crystal directions.

\subsection{Electronic Properties}
\begin{figure*}[htbp]
\centering
\includegraphics[scale=0.68]{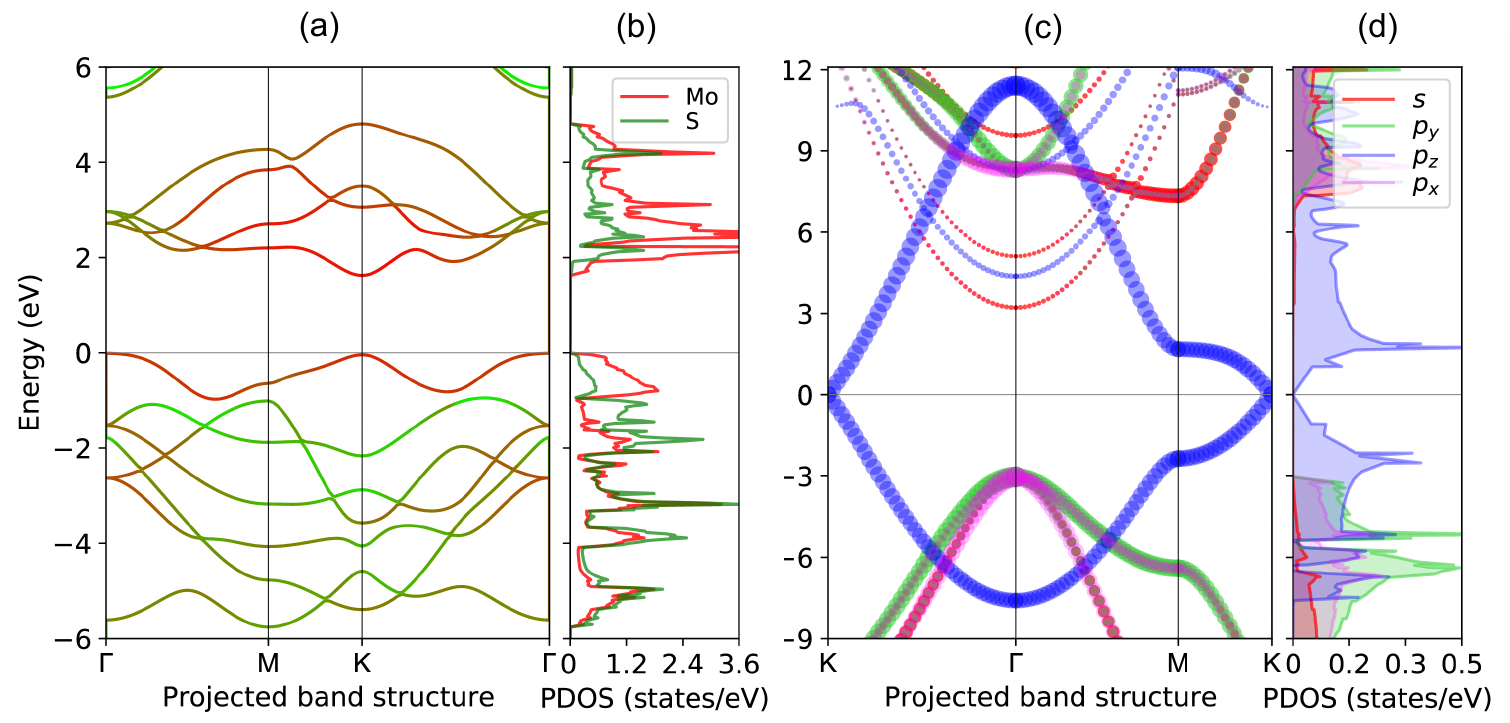}
\caption{\label{pband}(Color online) Projected band structure (left panel) and density of states (right panel) of (a) MoS$_2$ and (b) graphene monolayers. The Fermi energy is set to zero eV.}
\end{figure*}

\begin{figure*}[htbp]
\centering
\includegraphics[scale=0.23]{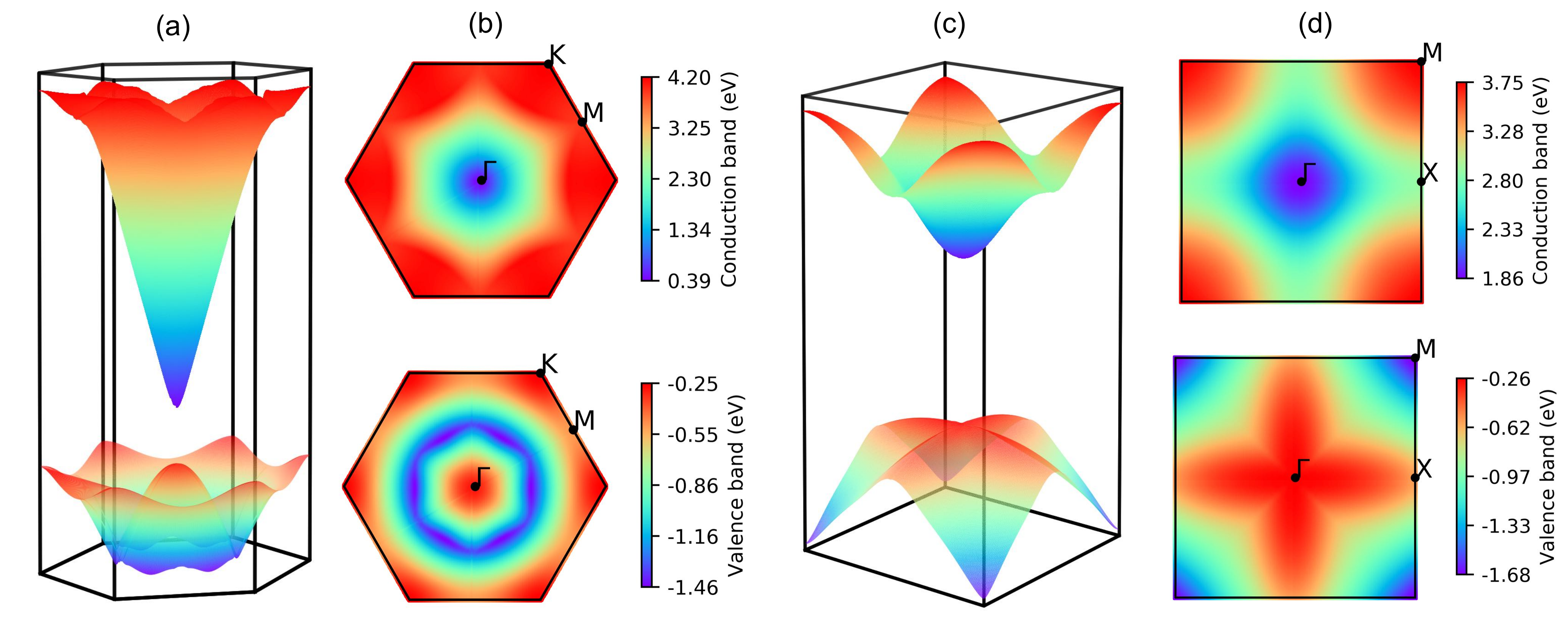}
\caption{\label{3D_band}(Color online) PBE calculated global band structure of (a) InN ($P\overline{6} m 2$) and (b) AgI (P4/nmm).}
\end{figure*}

\begin{figure*}[htbp]
\centering
\includegraphics[scale=0.55]{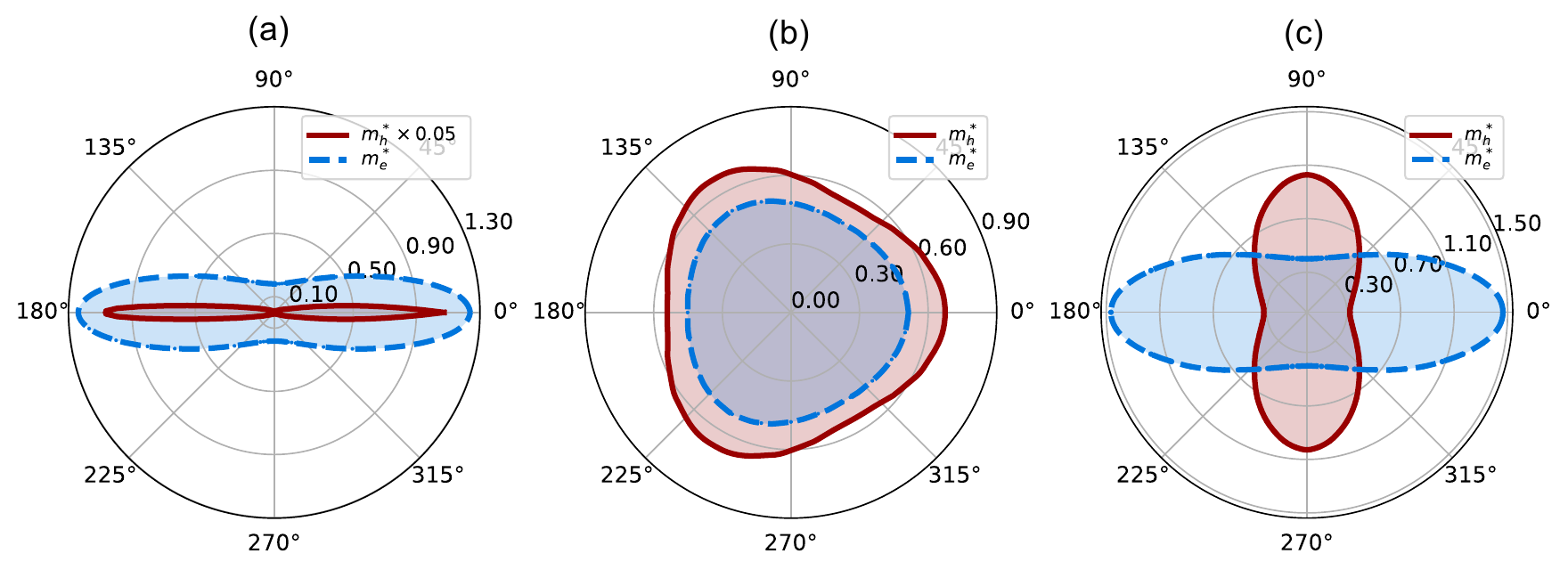}
\caption{\label{emc}(Color online) PBE calculated orientation-dependent effective masses  (in units of electron mass $m_0$) of  (a)  BP, (b) MoS$_2$ and (c) TiS$_3$ monolayers. The red and blue lines indicate the fitted  effective mass curves of hole and electron carriers, respectively.}
\end{figure*}

Beside the band structure, the projected band structure for each candidate  is also provided to illustrate the contributions of different atomic orbitals in energy and momentum space,  offering a chemist's perspective of the electronic structure. As examples, the element-resolved and orbital-projected band structures and the corresponding density of states (DOS) of MoS$_2$ and graphene monolayers are depicted in Fig. \ref{pband}.  To gain more insight into the topological characterization of band dispersions near Fermi energy, we calculated the global band structures of both VBM and CBM for each candidate at the PBE level. The global band structures of InN ($P\overline{6} m 2$) and AgI (P4/nmm) are illustrated in Fig. \ref{3D_band}. In addition, the orientation-dependent effective mass $m^*$($\theta$) of both holes and electrons can be further obtained from the global band structures,  with the aim of analyzing the anisotropic band dispersions. The PBE-calculated 2D polar representation curves for BP, MoS$_2$ and TiS$_3$ are presented in Fig. \ref{emc} for illustration purpose.   One can find that the effective masses of all representative semiconductors are highly anisotropic, especially for BP and TiS$_3$.  The calculated $m^*$ along $\Gamma$-X and $\Gamma$-Y  are  0.32 (1.52) $m_0$  and 1.06 (0.38) $m_0$ for hole (electron) in TiS$_3$ monolayer,  in good agreement with previous results, 0.32 (1.47) $m_0$ and 0.98 (0.41) $m_0$.\cite{Jin2015} By comparison, the effective mass of hole (electron) in MoS$_2$ slightly increases from 0.54 (0.44) $m_0$ along K-$\Gamma$ to 0.61 (0.47) $m_0$ along K-M  due to the higher hexagonal symmetry. We define the anisotropy ratios of effective masses, $\gamma_{h}$= $m_h^{max}$/$m_h^{min}$  for hole and  $\gamma_{e}$  =$m_e^{max}$/$m_e^{min}$ for electron carriers. The calculated $\gamma_{h}$ ($\gamma_{e}$) is 1.25 (1.14) for MoS$_2$,  3.18 (3.66) for TiS$_3$ and 128.67 (6.80) for BP. 

\begin{figure*}[htbp]
\centering
\includegraphics[scale=0.75]{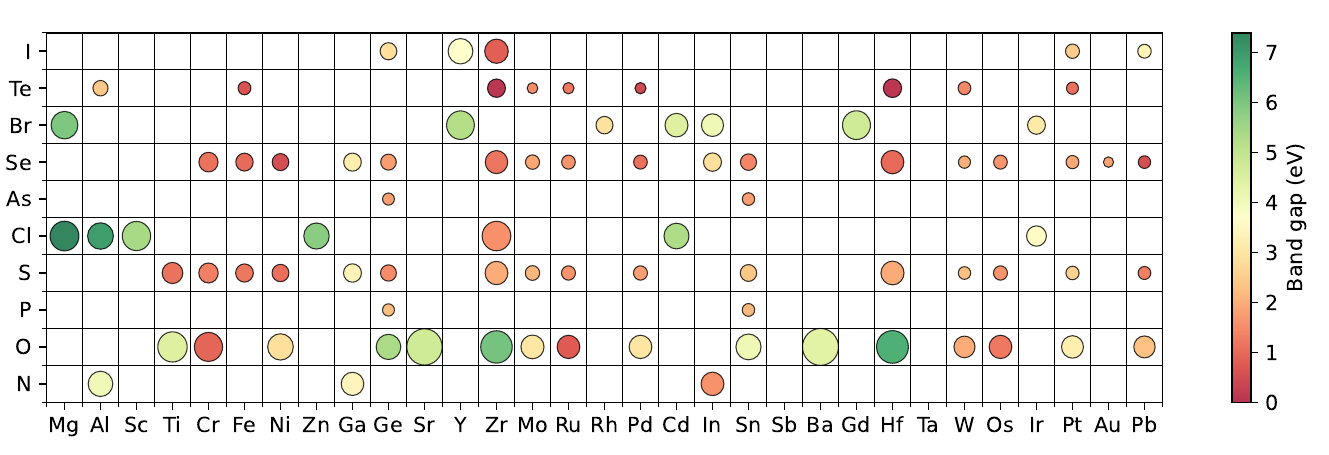}
\caption{\label{binary_gap}(Color online) HSE06 calculated band gap of binary 2D semiconductors as a function of  the electronegativity difference between two constituent elements. The circle radius indicates the electronegativity difference.}
\end{figure*}

To gain more insights into the band-gap variations of compounds, in Fig. \ref{binary_gap} we show the HSE06 predicted band gap ($E_g$) of binary 2D semiconductors as a function of the electronegativity difference between two constituent elements. The introduction of electronegativity difference here is to roughly evaluate the ionic character of the chemical bond formed between different elements. A larger difference in electronegativity implies a stronger ionic character. Overall, it is found that a compound with a stronger ionic bond tends to own a larger band gap value. Nevertheless, there are some exceptional cases in which small gaps come along with large electronegativity difference, such as CrO$_2$, ZrCl$_2$ and HfSe$_3$.
In the electronic and optoelectronic devices applications, not only the band gap, but also the absolute position of the band edges relative to vacuum, including ionization energy ($I$) and electron affinity ($A$) and work function ($\phi$) are important parameters.

\begin{figure}[htbp]
\centering
\includegraphics[scale=0.9]{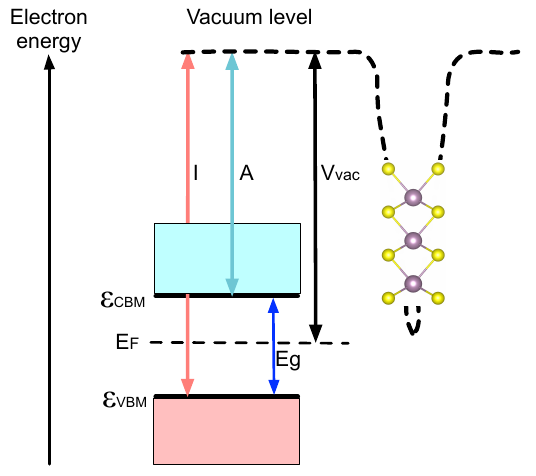}
\caption{\label{align}(Color online) Schematic energy diagram of a semiconductor. The ionization energy $I$, electron affinity $A$ and work function $\phi$ defined as the energies of VBM, CBM and Fermi level $E_F$ with respect to the vacuum level $V_\text{vac}$, respectively. }
\end{figure}

$I$ is the minimum energy needed to remove an electron from the highest occupied state to the vacuum, i.e. at $V_\text{vac}$, $I$=$V_\text{vac}-\varepsilon_\text{VBM}$. $A$ is the negative of the energy change when adding an electron to the lowest unoccupied state, $A$=$V_\text{vac}-\varepsilon_\text{CBM}$. Clearly,  the absolute positions VBM and CBM with respect to $V_\text{vac}$ are the negatives of $I$ and $A$, respectively.
The work function ($\phi$) is defined as the minimal energy needed to remove an electron originally at the Fermi level ($E_F$) deep inside the material to just outside its surface, namely, $\phi$=$V_\text{vac}-E_F$. In semiconductors, $\phi$ varies with the position of the $E_F$ because $E_F$ is strongly sensitive to the preparation condition of the sample in the measurement  which determines to a large extent  concentration of various intrinsic and extrinsic defects. Figure \ref{align} provides a schematic illustration of different quantities involved. In the KS-DFT scheme, the calculation of $V_\text{vac}$ is straightforward as it equals to the asymptotic value of the planar-averaged Hartree potential in the vacuum region, as illustrated in Fig. \ref{align}. The band edges of several widely studied 2D semiconductors, together with available theoretical data  in literature,  are listed in Table \ref{align_table}. One can find that the HSE06 calculated $E_g$, $I$ and $A$ of the representative systems are in good agreement with previous reports.\cite{Wang2015b,Watanabe2004,Mak2010,Jun2016}

\begin{table*}[htbp]
\begin{ruledtabular}
\caption{\label{align_table} HSE06-calculated band gap $E_g$, ionization energy $I$ and electron affinity $A$. For comparison purposes, the available theoretical values from the previous literature are also shown.} 
\begin{tabular}{c|cc|cc|cc|}
&\multicolumn{2}{c|}{Band gap (eV) }
&\multicolumn{2}{c|}{Ionization energy (eV)}
&\multicolumn{2}{c|}{Electron affinity (eV)}\\
Material & Our work &  Literature & Our work & Literature  & Our work & Literature     \\
\hline
BP & 1.57    & 1.52  [\onlinecite{Cai2014}]  & 5.46  & 5.43 [\onlinecite{Cai2014}] & 3.89 & 3.91 [\onlinecite{Cai2014}]     \\
BN & 5.71      & 5.68 [\onlinecite{Tony2016}] & 6.60  & 6.56 [\onlinecite{Tony2016}] & 0.89   &  0.88 [\onlinecite{Tony2016}]   \\
MoS$_2$  & 2.18    &  2.15 [\onlinecite{Tony2016}] & 6.38 & 6.33 [\onlinecite{Tony2016}] & 4.20   & 4.18 [\onlinecite{Tony2016}]  \\
WSe$_2$  & 2.04    &  1.98 [\onlinecite{Kang2013}] & 5.49 & 5.82 [\onlinecite{Kang2013}] & 3.45   & 3.84 [\onlinecite{Kang2013}]  \\
TiS$_3$  & 1.15    & 1.06 [\onlinecite{Dai2015}] & 5.87 & 5.34 [\onlinecite{Dai2015}] & 4.72   & 4.28 [\onlinecite{Dai2015}] \\
\end{tabular} 
\end{ruledtabular}
\end{table*}

\subsection{Optical Properties}
\begin{figure*}[htbp]
\centering
\includegraphics[scale=0.68]{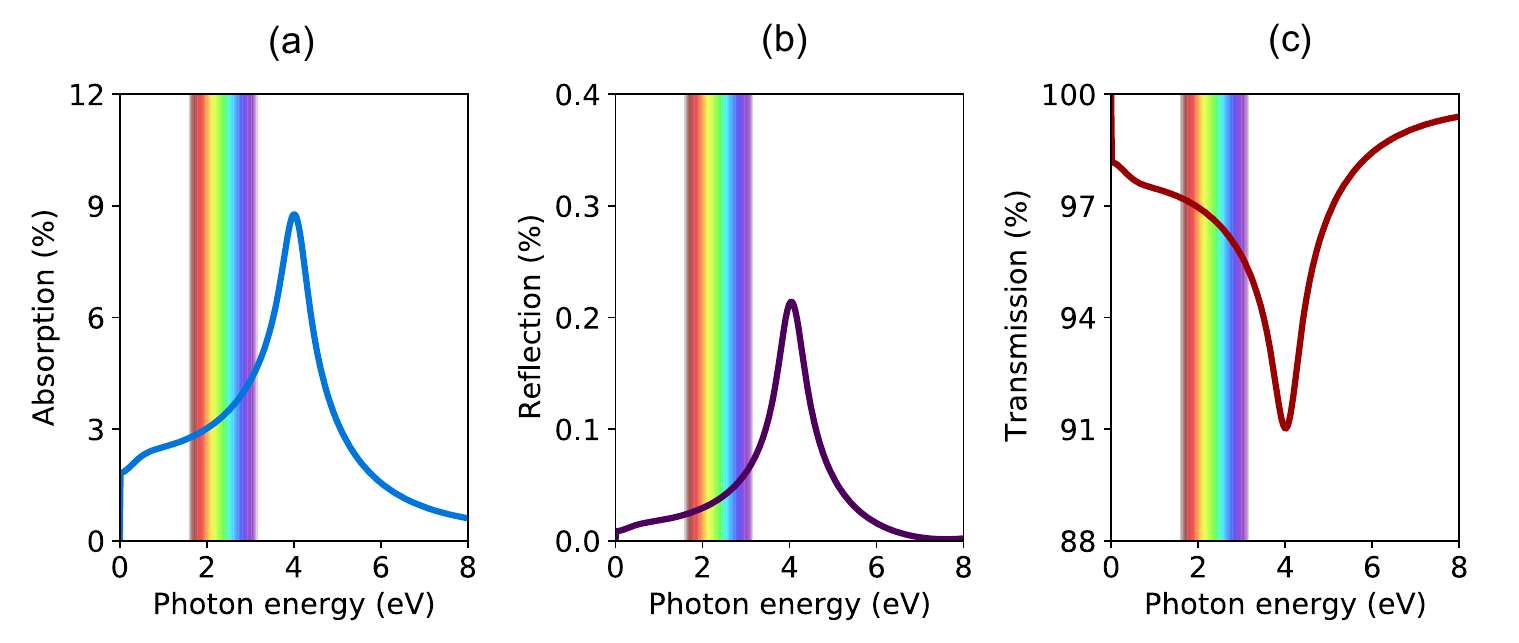}
\caption{\label{Optical_2D}(Color online) Frequency dependence of (a) absorbance, (b) reflectance and (c) transmittance for graphene.}
\end{figure*}

The macroscopic dielectric function of 2D materials cannot be well-defined with the layer thickness $d \rightarrow 0$. This is because the calculated dielectric function of an artificial 3D periodic system is affected by the length $L$ of the vacuum region in the standard DFT calculations. To avoid the thickness problem, an $L$-independent optical conductivity $\sigma_{2D}(\omega)$ is used to characterize the optical properties of 2D sheets,\cite{Matthes2014,Matthes2016} 

\begin{equation}
\sigma_{i j}(\omega)=\varepsilon_{0} \omega L\left[\varepsilon_{i j}(\omega)-\delta_{i j}\right],
\end{equation}

where $\varepsilon(\omega)$ is frequency-dependent complex dielectric function calculated in the framework of the independent-quasiparticle approximation\cite{Adolph1996}, $\varepsilon_{0}$ is the permittivity of vacuum, $\omega$ is the frequency of incident wave, and $L$ is the slab thickness in the simulation cell. In the present study we consider only the in-plane component $\varepsilon(\omega)$ of the dielectric tensor, $i.e.$, only light polarization perpendicular to the sheet normal has been taken into account. The normalized reflectance $R(\omega)$, the transmittance $T(\omega)$, and the absorbance $A(\omega)$ can be obtained from the following equation:\cite{Matthes2014,Matthes2016}

\begin{equation}\label{eq_op_2D}
\begin{aligned} R &=\left|\frac{\tilde{\sigma} / 2}{1+\tilde{\sigma} / 2}\right|^{2} \\ T &=\frac{1}{|1+\tilde{\sigma} / 2|^{2}} \\ A &=\frac{\operatorname{Re} \tilde{\sigma}}{|1+\tilde{\sigma} / 2|^{2}}  \end{aligned}
\end{equation}
where $\tilde{\sigma}(\omega)=\sigma_{2 \mathrm{D}}(\omega) / \varepsilon_{0} c$ is the normalized conductivity ($c$ is the speed of light). We present the linear optical properties of graphene in Fig. \ref{Optical_2D} as an illustrated example. 

\subsection{Scanning Tunneling Microscope Simulations} 
\begin{figure*}[htbp]
\centering
\includegraphics[scale=0.55]{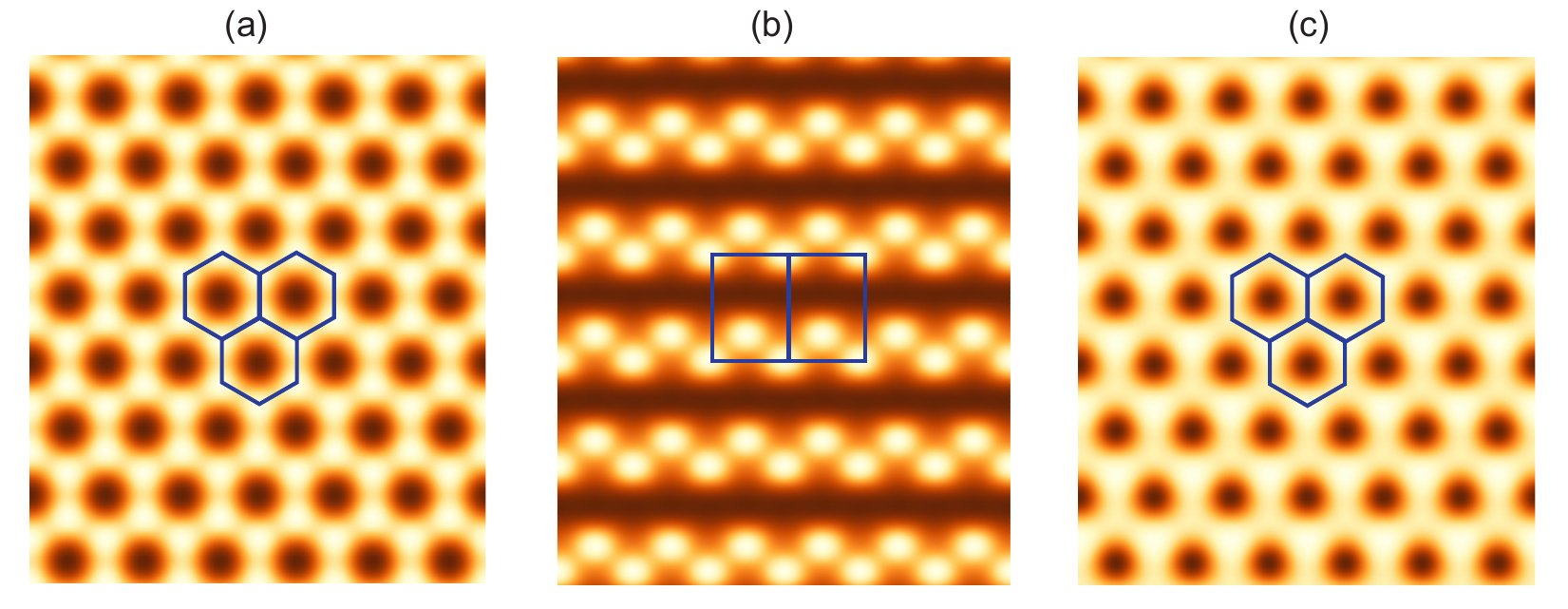}
\caption{\label{STM}(Color online) Simulated STM image of (a) graphene, (b) BP and (c) $h$-BN respectively.}
\end{figure*}

STM can not only characterize the atomic structure of material surfaces, but also can provide direct local insight into the electronic structure.\cite{Binnig1982} Thus, the simulated STM image has been obtained for each candidate based on the Tersoff-Hamann approach.\cite{Tersoff1985} In this model the calculated tunneling current $I$ which depends on the tip position $\mathbf{r}$ and the applied voltage $V$, is proportional to the integrated local density of states (LDOS)

\begin{equation}
I(\mathbf{r},V) \propto \int_{\epsilon_F}^{\epsilon_{F}+e V} \sum_{k n} w_{\mathbf{k}}\left|\Psi_{\mathbf{k} n}(\mathbf{r})\right|^{2} \delta\left(\epsilon-\epsilon_{\mathbf{k} n}\right) d \epsilon,
\end{equation}

where $V$ is the bias voltage, $w_{\mathbf{k}}$ is the $k$-point weight, $\Psi_{\mathbf{k} n}(\mathbf{r})$ and $\epsilon_{\mathbf{k} n}$  are the wave function and eigenvalue at the wave-vector $\mathbf{k} $ with band index $n$, and $\epsilon_{F}$ is the Fermi-energy. To simulate  STM images, we integrated the LDOS from 0.5 eV below the VBM up to 0.5 eV above the CBM. We chose the tunneling tip of 1.0 {\AA} and 2.0 {\AA} above the upper surface of 2D semiconductors during the simulations, respectively. Constant current topographs are approximated by constant charge density isosurfaces.  In Figs. \ref{STM}(a)-(c), we give the calculated STM images of graphene, BP and $h$-BN with examples. Clearly, we observe that the patterns in the computational and experimental STM images are very similar.\cite{Yu2011,Kiraly2017,Chen2020}

\subsection{ Linear Fitting Models of Band Edges}
\begin{figure}[htbp]
\centering
\includegraphics[scale=0.45]{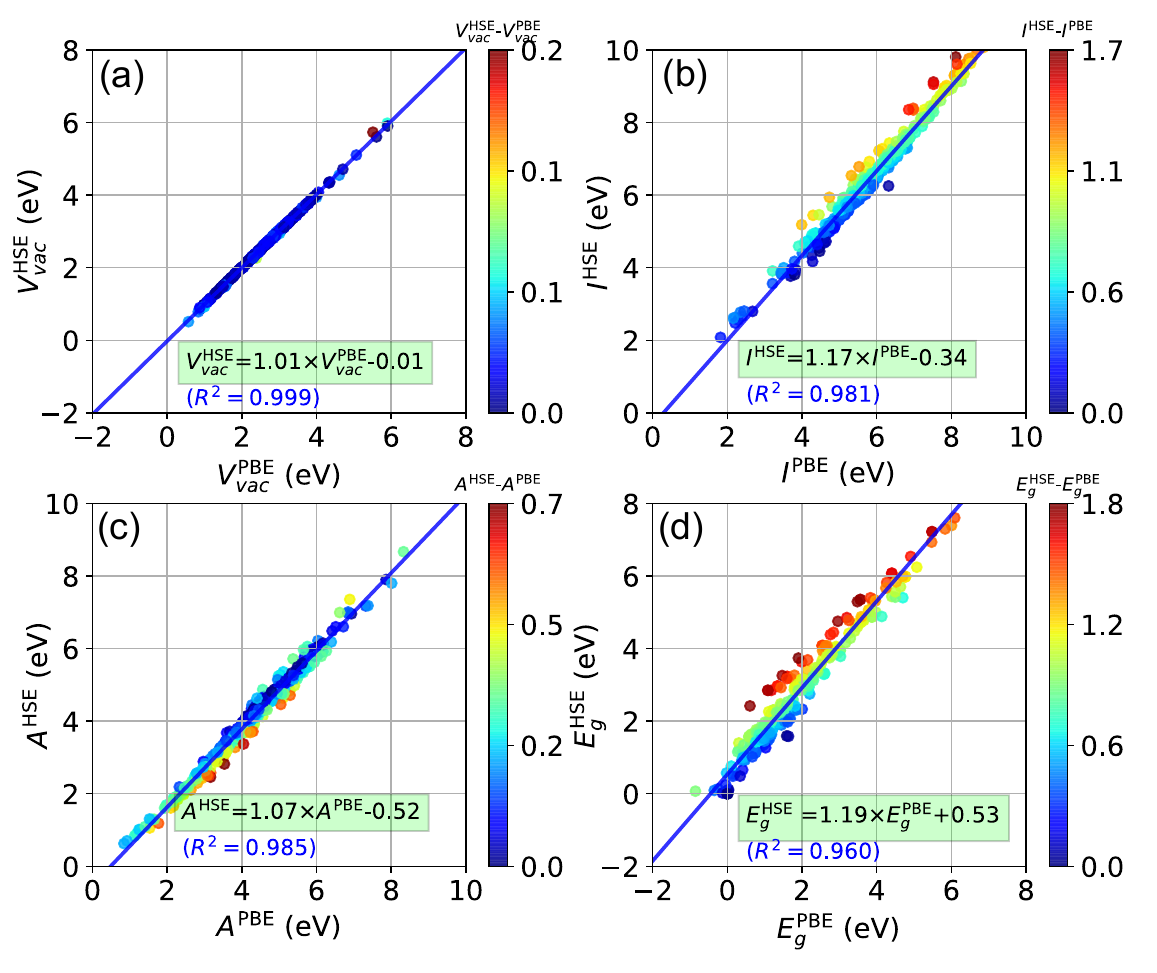}
\caption{\label{gap}(Color online) A comparison of PBE, HSE06 and LFM data of (a) vacuum level, (b) ionization energy, (c) electron affinity and (d) band gap, respectively. The color denotes the difference between HSE06 and LFM data. }
\end{figure}

It is well known that the conventional DFT calculations with local or semi-local exchange-correlation functionals underestimate the values of band gaps for most insulators and semiconductors, while the hybrid DFT makes better prediction, yet with very high computational cost. To gain additional insight into the difference between the PBE and HSE06 calculated band edges for 268 benchmark materials, we first plot the HSE06 calculated vacuum level ($V_\mathrm{vac}^\mathrm{HSE}$) versus the PBE result  ($V_\mathrm{vac}^\mathrm{PBE}$) in Fig. \ref{gap}(a). One can find that the vacuum level is independent of the exchange-correlation functional. Interestingly, although PBE systematically underestimates the ionization energy $I$, electron affinity $A$ and  band gap $E_g$ compared to HSE06, a  quasi-linear variation relation still hold true for PBE and HSE06 calculated band edges as shown in Figs. \ref{gap}(b)-(d). A linear least-squares fit gives 


\begin{equation}
\label{pred_V}
V_\mathrm{vac}^\mathrm{fit}=1.01 \times V_\mathrm{vac}^\mathrm{PBE}-0.01 \; (R^2 = 0.999), 
\end{equation}

\begin{equation}
\label{pred_I}
I^\mathrm{fit}=1.17\times I^\mathrm{PBE}-0.34 \; (R^2 = 0.981), 
\end{equation}

\begin{equation}
\label{pred_A}
A^\mathrm{fit}=1.07\times A^\mathrm{PBE}-0.52 \; (R^2 = 0.985), 
\end{equation}
and
\begin{equation}
\label{pred_E}
E_g^\mathrm{fit} =1.19\times E_g^\mathrm{PBE}+0.53 \; (R^2 = 0.960),
\end{equation}

where $R^2$ is the coefficient of determination indicating the proportion of data points which lie within the line created by the regression equation. Considering that the experimental data of the most of benchmarked systems are not available, we choose the HSE06 calculated data as a reference to evaluate the mean absolute error (MAE). The MAEs are 0.15 eV versus 0.64 eV for $I$, 0.12 eV versus 0.23 eV for $A$ and 0.25 eV versus 0.90 eV for $E_g$ obtained with LFM and PBE, respectively. Clearly, the linear fitting model (LFM) yields a drastically reduced MAE as compare to PBE, especially for the evaluation of $E_g$ and $I$. We further investigated the difference in band gap between HSE06 and LFM methods as a function of system as listed in the Supplemental Material), and find that the biggest deviation mainly occurs in the oxide semiconductors with heavy metal cations.  Although there are a wide class of materials which can be sufficiently well described by KS-DFT based on mean field theory approximation in which means an electron only experiences averaged out electrostatic interactions with other electrons, it fails to capture the physics of strongly correlated many-body effect.

\begin{table*}[htbp]
\begin{ruledtabular}
\caption{\label{align_table} Band gap $E_g$, ionization energy $I$ and electron affinity $A$ of typical few-layer semiconductors obtained with the PBE, HSE06 and LFM as described in Eqs. (\ref{pred_I})-(\ref{pred_E}).} 
\begin{tabular}{c|ccc|ccc|ccc|}
&\multicolumn{3}{c|}{Band gap (eV) }
&\multicolumn{3}{c|}{Ionization energy (eV)}
&\multicolumn{3}{c|}{Electron affinity (eV)}\\
Material			 &PBE 	& LFM	&  HSE06  &PBE & LFM &  HSE06   &PBE & LFM &  HSE06     \\
\hline
BP bilayer   				& 0.47		& 1.08  	& 1.08 	& 4.56 		& 5.00  & 4.95  	& 4.09 	& 3.86 	& 3.87 \\  
BP trilayer 				& 0.21		& 0.78  	& 0.80 	& 4.45 		& 4.86  & 4.82  	& 4.24 	& 4.01 	& 4.02 \\  
BP quadrilayer 			& 0.08		& 0.63  	& 0.66 	& 4.36 		& 4.76  & 4.72  	& 4.28 	& 4.06 	& 4.06 \\  
MoS$_2$ bilayer 		& 1.18		& 1.93  	& 1.77 	& 5.40 		& 5.98  & 5.87  	& 4.22 	& 4.00 	& 4.09 \\  
MoS$_2$ trilayer 		& 0.99		& 1.70  	& 1.57 	& 5.29 		& 5.85  & 5.74  	& 4.30 	& 4.08 	& 4.18 \\  
MoS$_2$ quadrilayer & 0.92		& 1.62 	 	& 1.49 	& 5.27 		& 5.82  & 5.71  	& 4.35 	& 4.13 	& 4.23 \\  
WS$_2$ bilayer  		& 1.35 		& 2.14 		& 1.94 	& 5.25 		& 5.81  & 5.68  	& 3.90 	& 3.66 	& 3.75 \\  
WS$_2$ trilayer			& 1.14 		& 1.89 		& 1.71 	& 5.13 		& 5.66  & 5.54  	& 3.99 	& 3.75 	& 3.84 \\  
WS$_2$ quadrilayer 	& 1.04 		& 1.76 		& 1.59 	& 5.07 		& 5.60  & 5.48  	& 4.04 	& 3.80 	& 3.89 \\  
BN bilayer 					& 4.42 		& 5.79 		& 5.75 	& 5.94 		& 6.61  & 6.73  	& 1.52 	& 1.11 	& 0.98 \\  
BN trilayer 				& 4.23 		& 5.56 		& 5.55 	& 5.89 		& 6.55  & 6.68  	& 1.67 	& 1.26 	& 1.13 \\  
BN quadrilayer 			& 4.18 		& 5.50 		& 5.51 	& 5.91 		& 6.57  & 6.71  	& 1.73 	& 1.33 	& 1.20 \\

\end{tabular} 
\end{ruledtabular}
\end{table*}

The above discussion provides us with important information about the assessment of band edges by the LFM with the PBE-calculated data as initial input. In order to benchmark the HSE06-calculated results with those obtained by using the LFM method as described in the Eqs. (\ref{pred_I})-(\ref{pred_E}), we revisited the band edges of some typical few-layer semiconductors using the PBE, HSE06 and LFM, respectively. We can see from Table \ref{align_table} that the agreement between the LFM and the corresponding HSE06 values is very good. The difference between these two data is less than 0.2 eV.

\begin{figure}[htbp]
\centering
\includegraphics[scale=0.28]{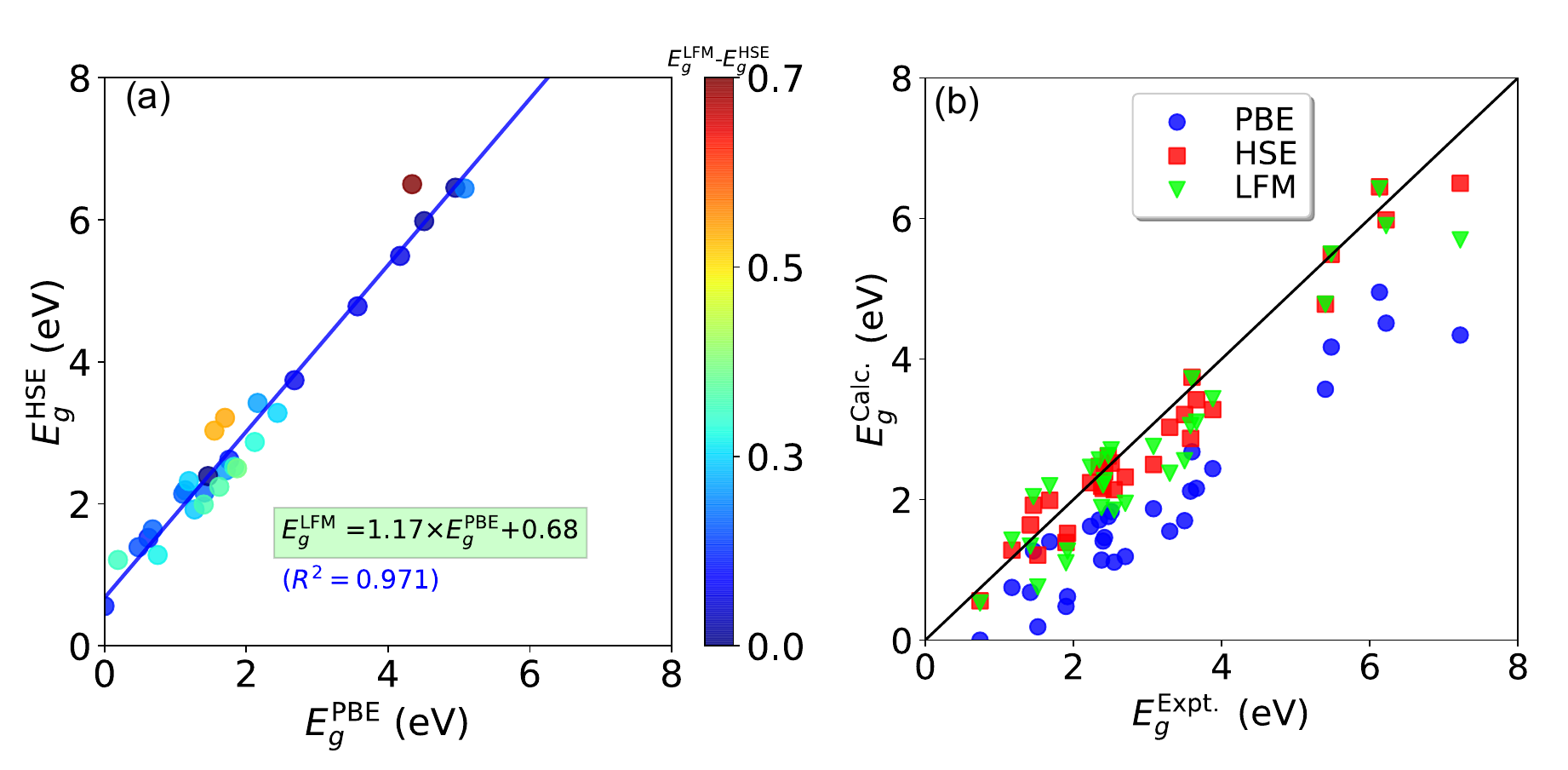}
\caption{\label{bulk}(Color online) A comparison of (a) HSE06 band gaps versus PBE ones and (b) experimental band gaps versus with values computed with PBE, HSE06 and LFM for the SC/40 set, respectively. The color denotes the difference between HSE06 and LFM data.}
\end{figure}

An interesting question arise: whether the LFM based on 2D semiconductors also holds true for bulk systems? We attempt to answer this question by examining the band gaps of the systems in the SC/40 test set.\cite{Heyd2005} In Fig. \ref{bulk}(a) we plot the HSE06 gaps versus the PBE ones of the SC/40 test set. A linear least-squares fit gives $E_g^\mathrm{fit} =1.17\times E_g^\mathrm{PBE}+0.68$ with R$^2$ = 0.971. One can find the data for both 2D and bulk systems exhibit a similar fit quality and interestingly also a similar slope of the fit lines even the size of bulk sample is not enough (around 40 materials). As shown in Fig. \ref{bulk}(b) the LFM predicted band gaps are comparable to those obtained with the HSE06 approach. Further calculations show that the and mean absolute relative error (MARE) with respect to the experimental data are 42.54\%, 18.41\% and 12.15\% for PBE, LFM and HSE06, respectively. Especially, for the $sp$ semiconductors without semicore $d$-electrons, the MAE between LFM and HSE06 data is only around 0.10 eV. This indicates that our LFM model can provide an accuracy similar to the HSE06 level but with low computational cost. On the other hand, when systems of the SC/40 test set includes cations with semicore delectrons, the MAE  can reach large to 0.42 eV.  

\section{Summary}
In conclusion, we have identified 259 2D nonmagnetic semiconductors from near 1000 2D monolayers by performing first-principles high-throughput calculations. The calculated properties include lattice constants, formation energy, Young's modulus, Poisson's ratio, scanning tunnel microscopy, band gap, band structure, anisotropic effective mass, ionization energy and electron affinity. We have also proposed a linear fitting model with a precision as high as hybrid DFT to evaluate band gap, ionization energy and electron affinity semiconductor from the PBE-calculated data as input. We expect that our computational screening database could stimulate further exploration of 2D semiconductors in nanoscale devices, and other important applications.
\section{Acknowledgement}
V.W. acknowledges the support of National Natural Science Foundation of China (Grant No. 62174136), Natural Science Basic Research Program of Shaanxi (Program Nos. 2022JQ-063 and 2021JQ-464), The Natural Science Basic Research Plan of Shaanxi Province (Grant No. 2021JZ-48), The Scientific Research Program Funded by Shaanxi Provincial Education Department (Grant Nos. 21JP088 and 22JP058), The Youth Innovation Team of Shaanxi Universities and Center for Computational Materials Science, Institute for Materials Research, Tohoku University for the use of MASAMUNE-IMR (Project No.2112SC0503). J.N. were supported by Innovative Science and Technology Initiative for Security Grant Number JPJ004596, ATLA, Japan.

\nocite{*}
\bibliographystyle{aipnum4-1}
%

\end{document}